\documentclass[aps,pra,showpacs,groupedaddress,superscriptaddress,twocolumn]{revtex4}

\usepackage{graphicx}
\usepackage{dcolumn}
\usepackage{bm}
\usepackage[usenames,dvipsnames]{color}
\usepackage{amsmath}
\usepackage[hypertex]{hyperref}
\usepackage{amsfonts}
\usepackage{amsthm}
\usepackage{amssymb}
\usepackage{mathrsfs}
\usepackage{subfigure}
\usepackage{ulem}

\begin{document}

\title{Non-Markovian Dynamics of Quantum Discord}

\author{F. F. Fanchini}
\email{fanchini@ifi.unicamp.br}
\affiliation{Instituto de F\'{\i}sica Gleb Wataghin, Universidade Estadual de
Campinas, P.O. Box 6165, 13083-970, Campinas, SP, Brazil}

\author{T. Werlang}
\affiliation{Departamento de F\'{\i}sica, Universidade Federal de S{\~a}o Carlos,
P.O. Box 676, 13565-905, S{\~a}o Carlos, SP, Brazil}

\author{C. A. Brasil}
\affiliation{Instituto de F\'{\i}sica de S\~{a}o Carlos, Universidade de S\~{a}o Paulo, P.O. Box 369, 13560-970, S\~{a}o Carlos, SP, Brazil}

\author{L. G. E. Arruda}
\affiliation{Instituto de F\'{\i}sica de S\~{a}o Carlos, Universidade de S\~{a}o Paulo, P.O. Box 369, 13560-970, S\~{a}o Carlos, SP, Brazil}

\author{A. O. Caldeira}
\affiliation{Instituto de F\'{\i}sica Gleb Wataghin, Universidade Estadual de
Campinas, P.O. Box 6165, 13083-970, Campinas, SP, Brazil}

\date{\today}

\begin{abstract}
We evaluate the quantum discord dynamics of two qubits in
independent and common non-Markovian environments. We compare the
dynamics of entanglement with that of quantum discord. For
independent reservoirs the quantum discord vanishes only at
discrete instants whereas the entanglement can disappear during a
finite time interval. For a common reservoir  quantum discord and
entanglement  can behave  very differently with sudden birth of
the former but not of the latter. Furthermore, in
this case the quantum discord dynamics presents sudden changes in
the derivative of its time evolution which is evidenced by the
presence of kinks in its behavior at discrete instants of time.
\end{abstract}

\pacs{03.65.Ud, 03.65.Yz, 03.67.Mn}

\maketitle
\section{INTRODUCTION}

Entanglement is a kind of quantum correlation (QC) that has been
playing a central role in quantum information and communication
theory \cite{nielsen}. However  there are other nonclassical
correlations apart from entanglement \cite{zurek,vedral,luo} that
can be of great importance to these fields. In order to
characterize all nonclassical correlations, Ollivier and Zurek
introduced what they called quantum discord \cite{zurek}. This
measure of quantum correlations captures a fundamental feature of
classical bipartite states - when the discord is zero the
information is locally accessible and can be obtained by distant
independent observers without perturbing the bipartite state.
Although a vast literature exists on the study of entanglement
just recently the other quantum correlations received due
attention \cite{luo,datta,white,phase
transition,Cui,lidar,werlang,jonas,cavalcanti}. A motivation for
the study of these correlations, for example, is the recent
discovery that  nonclassical correlations other than entanglement
can be responsible for the quantum computational efficiency of
deterministic quantum  computation with one pure qubit (DQC1)
\cite{datta,white}. In this context, the quantum discord could be
a new resource for quantum computation.

However, realistic quantum systems are not closed and therefore it
is of fundamental importance to study the quantum correlations
when the system loses its coherence due to interactions with the
environment \cite{breuer}. The entanglement dynamics in open
quantum systems was broadly studied in the literature but not much
exists about the effect of the environment on quantum discord
\cite{werlang,jonas,cavalcanti}. A peculiar aspect of the
entanglement dynamics is the well-known ``entanglement sudden
death'' (ESD) phenomenon \cite{eberly, yufirst}. This process
described the finite-time disentanglement of two parts that
interact with either independent \cite{yufirst,citeESD, bellomo07}
or common environments \cite{ficek, jppaz, mazzola}. In a previous
work \cite{werlang} we observed that, even at finite temperatures,
under a dissipative Markovian process, the quantum discord is
immune to ``sudden death''.

Despite the term ``entanglement sudden death''
sounds mysterious, it is important to note that there is no
dynamical distinction between separable and entangled states,
since the quantum states can, in general,  evolve back and forth
across the boundary between distinct full-dimensional subsets of
the space of the density matrices which contain separable and
entangled states.

The studies on the entanglement dynamics, that initially were
restricted to Markovian approximations, have recently been
extended to consider non-Markovian environments \cite{jppaz,
mazzola, bellomo07, nonM}. In this case, given the memory stored
in the environment, some of the initial entanglement that is lost
during the dissipative dynamics can return to the qubits. This
phenomenon is known as ``sudden-birth of entanglement''
(SBE), which in the light of what has been said
in the previous paragraph should not present much of a surprise to
us. Nevertheless a question still remains; what happens to
the quantum discord in this situation? Since quantum discord
exists even without entanglement, does it present sudden death
or even sudden birth? In order to answer these questions
we study the quantum discord of two qubits coupled to
non-Markovian dissipative environments.

In this paper we evaluate the quantum discord dynamics for a
dissipative non-Markovian process. For independent environments,
when the qubits are subject to amplitude damping,
we show that it only vanishes at discrete instants
of time, each within the time interval when the reduced quantum
state becomes pure and separable, and, consequently, the entanglement
vanishes. For a common reservoir the quantum discord behavior can
be very different from that of the entanglement. While the
entanglement dynamics presents damped oscillations with or without
sudden death, the quantum discord is almost always
positive and presents isolated kinks (cuspids) at which there is a
jump in its derivatives. The latter behavior is at clear variance
with what happens to the entanglement.

\section{QUANTUM DISCORD}

Entanglement is not the only measure of quantum correlations and
therefore an interesting approach was introduced in
\cite{zurek,vedral} to attempt to quantify all the nonclassical
correlations present in a system besides entanglement. The defined
quantity - the quantum discord - is given by the difference
between two expressions of mutual information (MI) extended from
classical to quantum system.

The total correlation between two classical systems $\mathcal{A}$
and $\mathcal{B}$, whose state is described by a joint probability
distribution $p\left(\mathcal{A},\mathcal{B}\right)$, can be
obtained by a measure of the MI,
${\mathcal{I}}\left(\mathcal{A}:\mathcal{B}\right)=H\left(\mathcal{A}\right)+H\left(\mathcal{B}\right)-H\left(\mathcal{A},\mathcal{B}\right)$,
where $H\left(\mathcal{\cdot}\right)$ denotes the Shannon entropy
$H\left(p\right)=-\sum_{jk}p_{jk}\log_2 p_{jk}$ \cite{nielsen}.
This classical MI can be rewritten as the equivalent expression
${\mathcal{J}}\left(\mathcal{A}:\mathcal{B}\right)=H\left(\mathcal{A}\right)-H\left(\mathcal{A}\mid\mathcal{B}\right)$
through the Bayes rule \cite{bayes}, where the conditional entropy
$H\left(\mathcal{A}\mid\mathcal{B}\right)$ quantifies the
ignorance about the state of $\mathcal{A}$ when one knows the
state of $\mathcal{B}$. For a quantum system represented by a
bipartite density operator $\rho$, the Shannon entropy functional
is replaced by the von Neumann entropy,
$S\left(\rho\right)=-\mathrm{Tr}\left(\rho\log_2\rho\right)$,
which is the first quantum extension of the classical MI. We
denote it ${\mathcal{I}}\left(\rho\right)$.

Another route to generalizing the classical MI to the quantum case
is to use a measurement-based conditional density operator
\cite{zurek}. If we restrict ourselves to projective measurements
performed locally only on system $\mathcal{B}$ described by a
complete set of orthogonal projectors, $\left\{ \Pi_{k}\right\} $,
corresponding to outcomes $k$, the quantum state\ after a
measurement changes to
$\rho_{k}=\left[\left(\mathbb{I}\otimes\Pi_{k}\right)\rho\left(\mathbb{I}\otimes\Pi_{k}\right)\right]/\mathrm{Tr}\left(\mathbb{I}\otimes\Pi_{k}\right)\rho\left(\mathbb{I}\otimes\Pi_{k}\right)\,$,
where $\mathbb{I}$ is the identity operator for system
$\mathcal{A}$. With this conditional density operator, a quantum
analogue of the conditional entropy can then be defined as
$S\left(\rho\mid\left\{ \Pi_{k}\right\}
\right)=\sum_{k}p_{k}S\left(\rho_{k}\right)$, and the second
quantum extension of the classical MI may be found,
$\mathcal{J}\left(\rho\mid\left\{ \Pi_{k}\right\}
\right)=S\left(\rho^{\mathcal{A}}\right)-S\left(\rho\mid\left\{
\Pi_{k}\right\} \right)$. Projective measurements on system
$\mathcal{B}$ remove all nonclassical correlations between
$\mathcal{A}$ and $\mathcal{B}$, but the value of
$\mathcal{J}\left(\rho\mid\left\{ \Pi_{k}\right\} \right)$ depends
on the choice of $\left\{ \Pi_{k}\right\} $. Therefore, to ensure
that it captures all classical correlations, we need to maximize
$\mathcal{J}$ over all $\left\{ \Pi_{k}\right\} $. This quantity,
$\mathcal{Q}\left(\rho\right)=\sup_{\left\{ \Pi_{k}\right\}
}\mathcal{J}\left(\rho\mid\left\{ \Pi_{k}\right\} \right)$, is
interpreted explicitly by Henderson and Vedral \cite{vedral}, as a
measure of classical correlations. The quantum discord is then
defined as \begin{eqnarray}\label{discord}
D\left(\rho\right)={\mathcal{I}}\left(\rho\right)-\mathcal{Q}\left(\rho\right),\end{eqnarray}
and provide us with information on the quantum nature of the
correlations between two systems, such that it is zero only for
states with classical correlations \cite{zurek,vedral} and nonzero
for states with quantum correlations. Although quantum discord is
equal to the entanglement of formation for pure states , it is not
true for mixed states, since some states present finite quantum
discord even without entanglement \cite{zurek}.

It is important to note that to calculate the classical
correlations one can consider arbitrary POVM measurements as
Henderson and Vedral did in \cite{vedral}. However, for two qubits, which is
our case, Hamieh \textit{et al.} \cite{ham} show that the projective measurement
is the POVM which maximizes the classical correlations.

\subsection{Analytical expression for quantum discord}
To evaluate the quantum discord dynamics presented in this
article we determine an analytical expression for a subclass of
the $X$ structured density operator. We consider a density matrix
as given by
\begin{equation}\label{matXd}
\rho(t)=\left(\begin{array}{cccc}
a & 0 & 0 & w\\
0 & b & z & 0\\
0 & z & b & 0\\
w & 0 & 0 & d\
\end{array}\right).
\end{equation}
where the coherences are real numbers and the element $\rho_{22}=\rho_{33}$.
It is easy to see that for this expression of $\rho(t)$ the
condition $S(\rho ^{\mathcal{A}})=S(\rho^{\mathcal{B}})$ is
satisfied  and therefore the measurement of classical correlations
assumes equal values, irrespective of whether the measurement is
performed on the subsystem $\mathcal{A}$ or $\mathcal{B}$
\cite{vedral}. To reduce the difficulty to compute  the quantum
discord we need to be able to maximize the classical correlation
$\mathcal{Q}(\rho)$. This can be done analytically if one notes
that a general one-qubit projector can be written as a function of
two angles, since
\begin{equation}
\mathcal{Q}(\rho)=S(\rho^\mathcal{A})-F(\theta,\phi),
\end{equation}
where
\begin{equation}
F(\theta,\phi)=\inf_{\{\theta,\phi\}}\left[\sum_{k=1,2}p_k(\theta,\phi)
S\left(\frac{\Pi_k^\mathcal{B}(\theta,\phi)\rho^\mathcal{AB}\Pi_k^\mathcal{B}
(\theta,\phi)}{p_k(\theta,\phi)} \right)\right],
\end{equation}
with $p_k(\theta,\phi)=\mbox{Tr}\left\{\Pi_k^\mathcal{B}(\theta,\phi)
\rho^\mathcal{AB}\Pi_k^\mathcal{B}(\theta,\phi)\right\}$ and the projectors
$\Pi_k^\mathcal{B}(\theta,\phi)=\mathbb{I}\otimes\left|k\right\rangle\left\langle k\right|$ $\left(k=1,2\right)$
defined by the orthogonal states:
\begin{eqnarray}
\left\vert 1\right\rangle &=&\cos{\theta }\left\vert \uparrow\right\rangle +e^{i\phi }\sin {\theta }\left\vert\downarrow\right\rangle,\nonumber\\
\left\vert 2\right\rangle &=&\sin{\theta }\left\vert \uparrow\right\rangle -e^{i\phi }\cos {\theta }\left\vert\downarrow\right\rangle.
\end{eqnarray}

We begin noting some peculiar properties of $F(\theta,\phi)$ when
$\rho^{\mathcal{AB}}$ is given by Eq. (\ref{matXd}). Given the
structure of the density matrix the critical points of
$F(\theta,\phi)$, i.e. the set of values of $\theta$ and $\phi$
 such that $\frac{\partial F(\theta,\phi)} {\partial\theta}=0{\rm{\;\; and
\;\;}} \frac{\partial F(\theta,\phi)}{\partial\phi}=0$,
do not depend on the elements of the density matrix. 
For $\theta=n\frac{\pi}{2}$ with $n\in \mathbb Z$ we have a set of
critical points and in this case the function $F(\theta,\phi)$
does not depend on the angle $\phi$. Another set is given by
$\theta=m\frac{\pi}{4}$ and $\phi=n\frac{\pi}{2}$ with $m,n \in
\mathbb Z$. Thus, with this observation, using the quantum version
of the mutual information $\mathcal{I}$ and Eq. (\ref{discord}),
it is straightforward to compute an analytical expression for the
quantum discord:
\begin{eqnarray}
D\left(\rho\right)=\min\left\{D_1,D_2\right\}
\end{eqnarray}
where
\begin{eqnarray}\label{d1}
D_1&=&S(\rho^\mathcal{A})-S(\rho^{\mathcal{A}\mathcal{B}}) - a\log_2\left(\frac{a}{a+b}\right)-b\log_2\left(\frac{b}{a+b}\right)\nonumber\\
   &-&d\log_2\left(\frac{d}{b+d}\right)-b\log_2\left(\frac{b}{d+b}\right),
\end{eqnarray}
and
\begin{eqnarray}\label{d2}
\hspace{-0.5cm}D_2&=&S(\rho^\mathcal{A})-S(\rho^{\mathcal{A}\mathcal{B}})-\Delta_+\log_2\Delta_+ -\Delta_-\log_2\Delta_-,
\end{eqnarray}
with $\Delta_\pm=\frac{1}{2}\left(1\pm\Gamma\right)$ and $\Gamma^{2} = \left(a-d\right)^{2}+4\left(|z|+|w|\right)^{2}$, which has been numerically verified for any density operator with the same structure as in Eq. (\ref{matXd}).

\section{EXACT DISSIPATIVE DYNAMICS OF QUANTUM DISCORD}
In this article we study a system whose dynamics is described by
the well-known damped Jaynes-Cummings model. We consider two
distinct situations, the independent and common environment. In
the former, each qubit is coupled to its own reservoir since the
dissipative processes occur independently. In the common
environment case, on the other hand, we consider only one bath for
both qubits. We suppose that the qubits are coupled to a single
cavity mode which in turn is coupled to a non-Markovian
environment that initially is in the vacuum state. In this case we
can say that the interaction Hamiltonian reduces the amplitude of
motion of each qubit state, and this justifies why such a process
is known in the literature as the amplitude damping channel. The
solutions of these simple models have recently been used to study
the non-Markovian effects on the dynamics of entanglement
\cite{bellomo07,mazzola}. The environments are represented by a
bath of harmonic oscillators, and the spectral density is of the
form
\begin{eqnarray}
J(\omega)=\frac{1}{2\pi}\frac{\gamma_{0}\lambda^{2}}{(\omega_{0}-\omega)^{2}+\lambda^{2}},\label{jw}\end{eqnarray}
 where $\lambda$ is connected to the reservoir correlation time $\tau_{B}$
by the relation $\tau_{B}\approx1/\lambda$, and $\gamma_{0}$ is
related to the time scale $\tau_{R}$ over which the state of the
system changes, $\tau_{R}\approx1/\gamma_{0}$. Here we will consider the strong coupling limit, i. e. $\tau_R<2\tau_B $.

For independent amplitude damping channels the two-qubit
Hamiltonian can be written as
\begin{equation}
H=\omega^{(i)}_{0}\sigma^{(i)}_{+}\sigma^{(i)}_{-}+\sum_k\omega^{(i)}_{k}{a^{(i)}_{k}}^{\dagger}a^{(i)}_{k}+\left(\sigma^{(i)}_{+}B^{(i)}+\sigma^{(i)}_{-}{B^{(i)}}^{\dagger}\right),
\end{equation}
 where $B^{(i)}=\sum_{k}g^{(i)}_{k}a^{(i)}_{k}$ with $g^{(i)}_{k}$ being the coupling
constant, $\omega^{(i)}_{0}$ is the transition frequency of the
i-$th$ qubit, and $\sigma^{(i)}_{\pm}$ are the system raising and
lowering operators of the i-$th$ qubit. Here the index $k$ labels
the reservoir field modes with frequencies $\omega^{(i)}_{k}$, and
${a^{(i)}_{k}}^{\dagger}$ ($a^{(i)}_{k}$) is  their usual creation
(annihilation) operator. Here and in the following the Einstein
convention sum is adopted.

For common environments, on the  other hand, we have that the
system raising (lowering) operator of each qubit is coupled to the
same environment operator $B$ $(B^\dagger)$. In this case we have
one bath coupled to both qubits, and the Hamiltonian is given by
\begin{equation}
H=\omega^{(i)}_{0}\sigma^{(i)}_{+}\sigma^{(i)}_{-}+\sum_k \omega_{k}{a_{k}}^{\dagger}a_{k}+\left(\sigma^{(i)}_{+}B+\sigma^{(i)}_{-}{B}^{\dagger}\right).
\end{equation}
We have considered two identical atoms equally coupled to the
reservoir. In this case, the dynamics of the two qubits occur in
two completely decoupled subspaces, generated by
$\left\{\left|00\right\rangle,\left|+\right\rangle=\left(\left|10\right\rangle
+ \left|01\right\rangle\right)/\sqrt{2},\left|11\right\rangle
\right\}$ and
$\left\{\left|-\right\rangle=\left(\left|10\right\rangle -
\left|01\right\rangle\right)/\sqrt{2}\right\}$. Using this fact,
Mazzola \textit{et al.} \cite{mazzola} connect the problem with a
three-level ladder system \cite{3level} and through the pseudomode
approach \cite{pseudomode}, the dynamics can be evaluated without
any approximations.

The initial state considered in this paper is the Bell-like state:
\begin{eqnarray}\label{istate}
\left|\psi\right\rangle = \alpha\left|00\right\rangle + \sqrt{1-\alpha^2}\left|11\right\rangle,
\end{eqnarray}
and thus the density matrix of the atomic system has the form of
Eq. (\ref{matXd}) with the dynamics of its matrix elements  given
in \cite{bellomo07} for independent reservoir and
\cite{mazzola,mazzola01} for common reservoir.

\begin{figure}[htbp]
\begin{center}
\includegraphics[width=.48\textwidth]{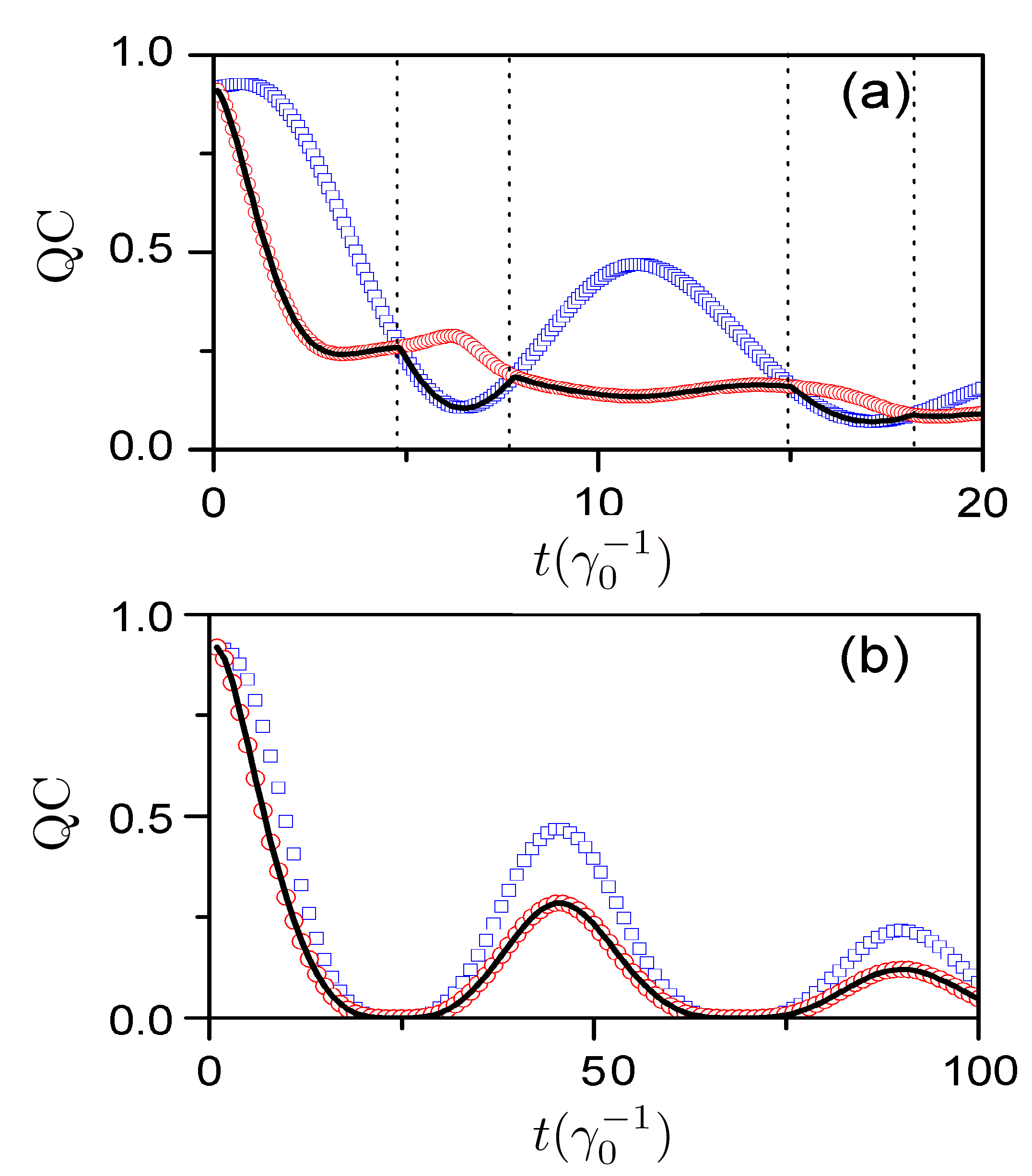} {}
\end{center}
\caption{(Color online) Analytical ($D_1$-squares, and $D_2$-circles) and
numerical (solid line) dynamics of discord for (a) a common reservoir with
$\lambda=0.1\gamma_0$ and (b) independent reservoirs with
$\lambda=0.01\gamma_0$. The dotted lines in (a) indicate the value
of $\gamma_0t$ where the sudden change occurs. The two-qubits
initial state used here is given in Eq. (\ref{istate}) with
$\alpha^2=1/3$.}\label{fig1}
\end{figure}

In Fig. (\ref{fig1}) we plot the quantum discord as a function of
the scaled time $\gamma_0t$ in the strong coupling regime,
$\lambda=0.1\gamma_0$  for the common reservoir case and
$\lambda=0.01\gamma_0$ for independent reservoirs. The initial
state used in Fig. (\ref{fig1}) is given by Eq. (\ref{istate})
with $\alpha^2=1/3$. The analytical solution of the quantum
discord is the minimum value assumed by  the functions
$D_1$(squares) and $D_2$(circles), given by Eq. (\ref{d1}) and Eq.
(\ref{d2}) respectively. In Fig. (\ref{fig1}) it is represented by
the black (solid) line.

We have observed that, for the values of $\gamma_0t$ indicated by
the doted lines in Fig (\ref{fig1}), the angles that minimize the
discord change, point out to a ``sudden change'' of discord as
previously noted in \cite{jonas} for Markovian environments. In
the region between two consecutive dotted lines the angles remain
unchanged. As shown in Fig.1b, the same analysis applies to
independent reservoirs, but no ``sudden change'' is observed in
this case. Here it should be emphasized that what
we are calling ``sudden change of discord'' is actually a
signature of a jump of the time derivatives of that function at
specific instants.

\begin{figure}[!htbp]
\begin{center}
\includegraphics[width=.48\textwidth]{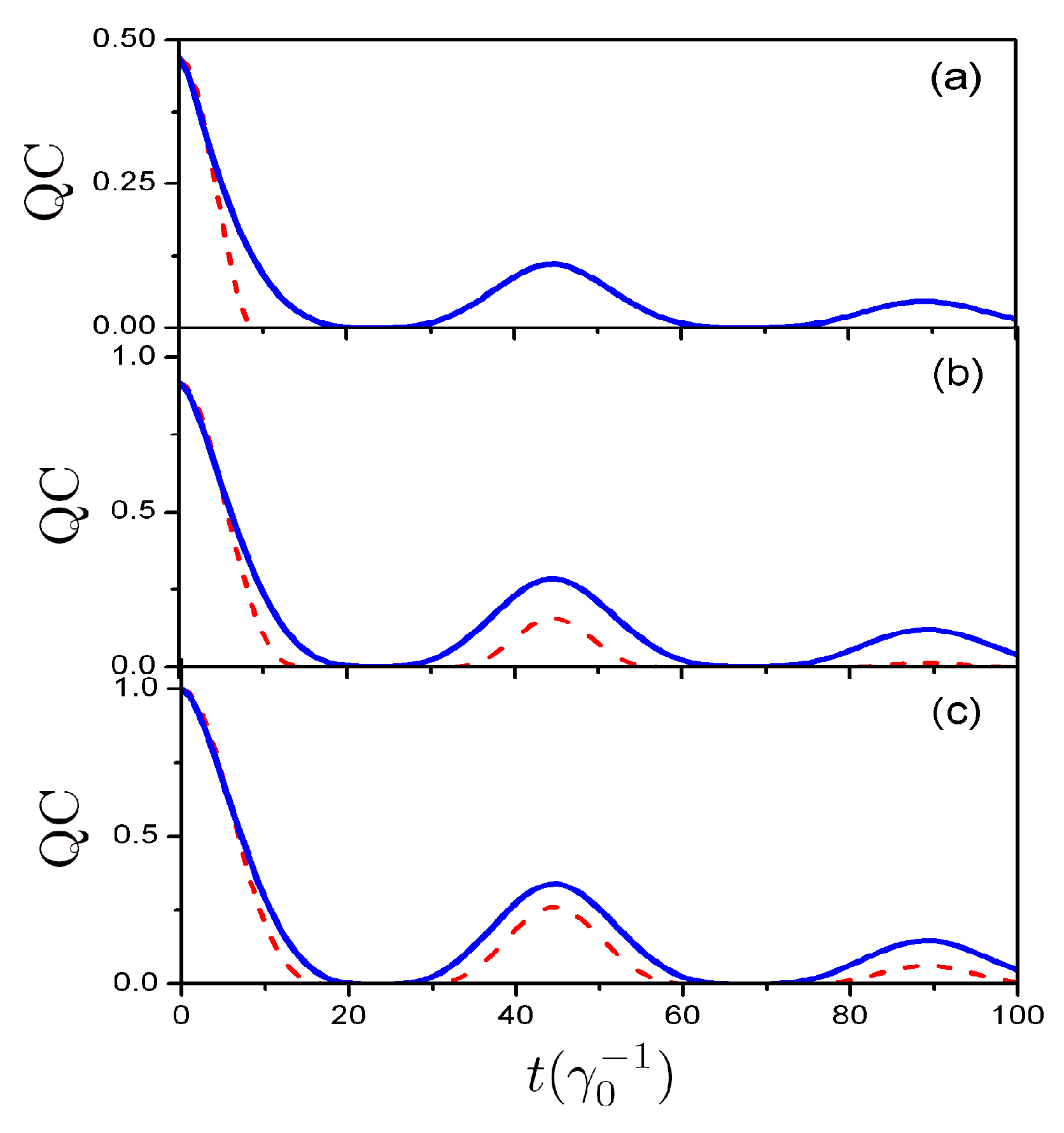} {}
\end{center}
\caption{(Color online) Dynamics of Discord (solid line) and EoF (dashed line) as
a function of the scaled time for independent reservoirs with
$\lambda=0.01\gamma_0$ and the two qubits initially prepared in
the state of eq. (\ref{istate}) with (a) $\alpha^2=1/10$, (b)
$\alpha^2=1/3$, and (c) $\alpha^2=1/2$.}
\end{figure}

\begin{figure}[!htbp]
\begin{center}
\includegraphics[width=.45\textwidth]{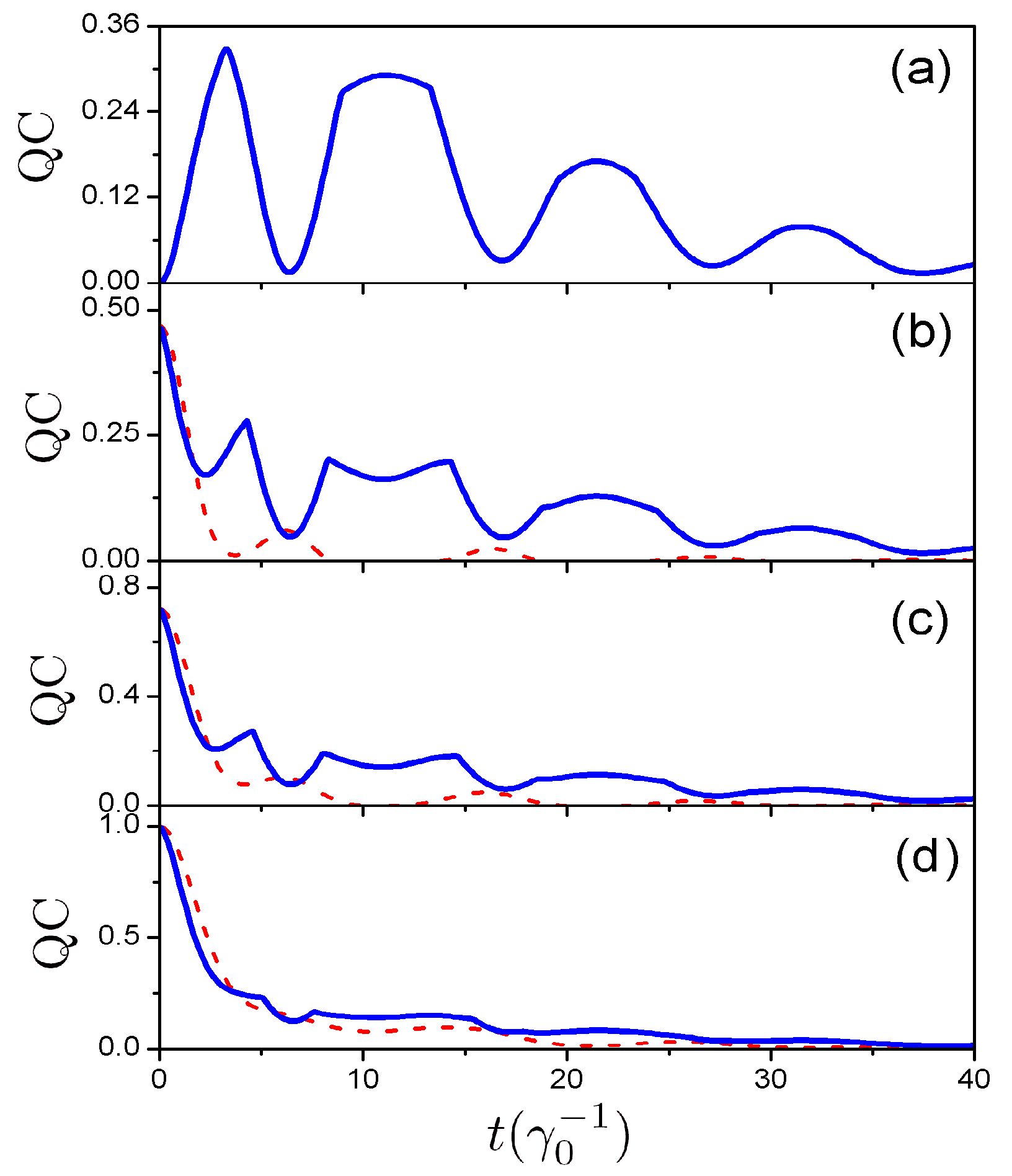} {}
\end{center}
\caption{(Color online) Dynamics of Discord (solid line) and EoF (dashed line) as
a function of the scaled time for common reservoirs with
$\lambda=0.1\gamma_0$ and two qubits initially prepared in the
state of eq. (\ref{istate}) with (a) $\alpha^2=0$, (b)
$\alpha^2=1/10$, (c) $\alpha^2=1/5$, and (d) $\alpha^2=1/2$.}
\end{figure}

\begin{figure}[!htbp]
\begin{center}
\includegraphics[width=.48\textwidth]{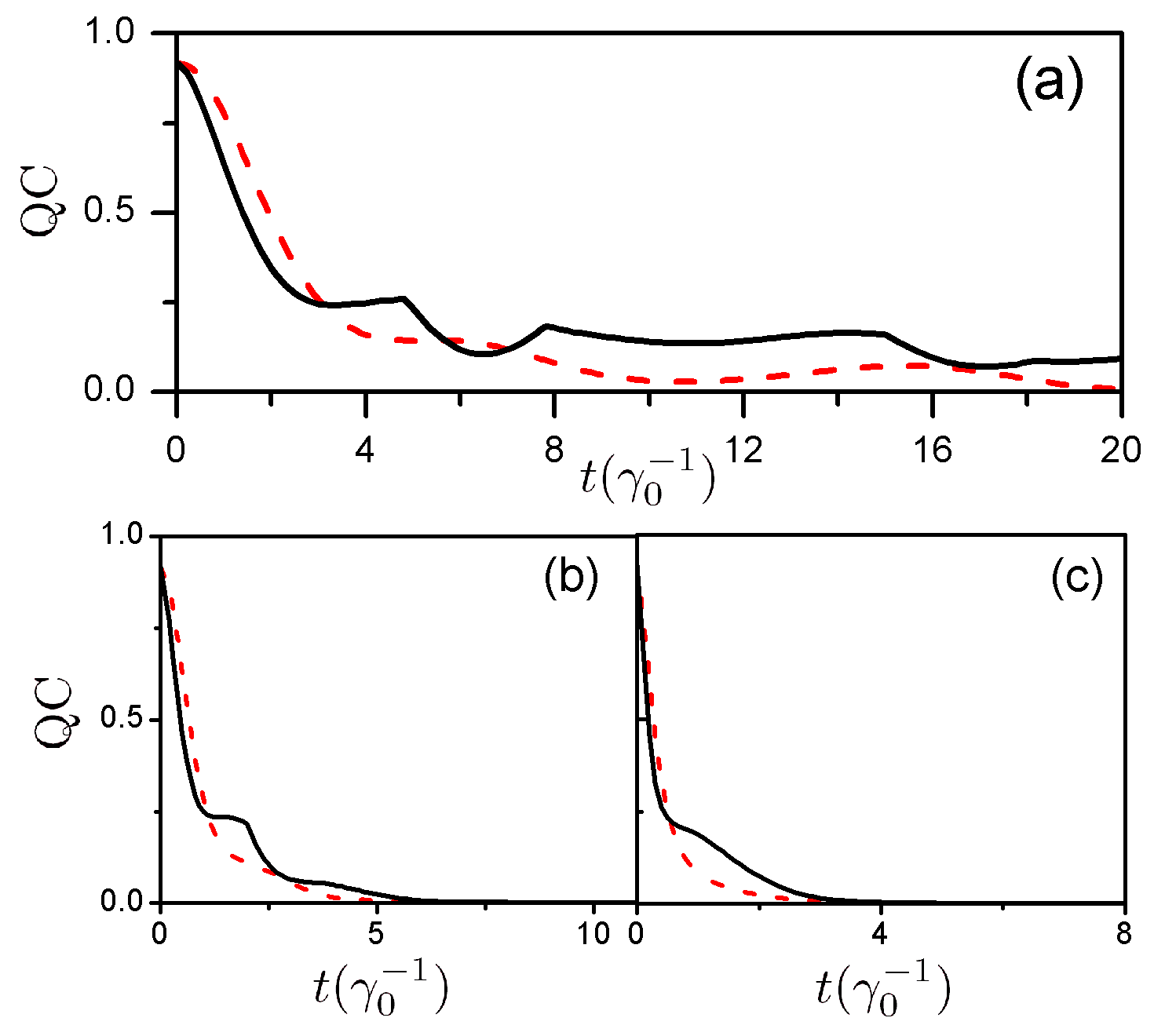} {}
\end{center}
\caption{(Color online) Dynamics of Discord (solid line) and EoF (dashed line) as
a function of the scaled time for a common reservoir with (a)
$\lambda=0.1\gamma_0$, (b) $\lambda=\gamma_0$, and (c)
$\lambda=10\gamma_0$. The two qubits initial state is given by eq.
(\ref{istate}) with $\alpha^2=1/3$.}
\end{figure}

In order to compare the discord dynamics with the entanglement
dynamics we used the entanglement of formation (EoF) \cite{eof} as
a measure of entanglement. For two qubits, the EoF dynamics can be
written as a function of the concurrence \cite{concurrence} and it
is  given by
\begin{equation}
E(t) = -\Gamma(t)\log_2\Gamma(t) - [1-\Gamma(t)]\log_2[1-\Gamma(t)]
\end{equation}
where $\Gamma(t)=\frac{1}{2}(1 + \sqrt{1 - C(t)^2})$ with $C(t)$
being the time dependent concurrence. For a density matrix with a
structure defined as in Eq. (\ref{matXd}) we have that
$C(t)=2\rm{max}\{0,\Lambda_1(t),\Lambda_2(t)\}$ with
$\Lambda_1(t)=z(t) - \sqrt{a(t)d(t)}$ and $\Lambda_2(t)=w(t) -
b(t)$.

We begin analyzing the entanglement and quantum discord dynamics
for independent reservoirs. The entanglement dynamics, for
example,  shows different behaviors depending on the initial state
of the two-qubit system \cite{bellomo07}. In the case where
$\alpha^2\geq1/2$ the EoF periodically vanishes for the discrete
times defined by $t_n=2\left[n\pi-\arctan{d/\lambda}\right]/d$
with $d=\sqrt{2\gamma_0\lambda-\lambda^2}$ and $n$ an integer
\cite{bellomo07}. This behavior is illustrated in Fig. (2c)
(dashed line) where we observe that the amplitude of oscillation
undergoes a decay after each revival. However, when $\alpha^2>1/2$
the behavior of the EoF presents two different features: $(i)$
there is ESD because the EoF  permanently vanishes within  finite
time intervals [see Fig.2a (dashed line)] and $(ii)$ the revival
of entanglement after these intervals when the two qubits are
fully disentangled [see Fig.2b (dashed line)].

The discord dynamics for all initial entangled  states
($0<\alpha^2<1$) is similar to the EoF for $\alpha^2\geq1/2$,
i.e., the discord vanishes only at $t_n$ when the two-qubits state
becomes the separable pure  state $\left|00\right\rangle$ [see
Fig.2a-c (solid line)]. The nonclassical correlations are mediated
by the reservoir, since there is no interaction between the
qubits. Furthermore, whereas the entanglement may reappear after a
time interval within which the EoF is zero, the discord is almost
always non zero. This result indicates that the discord under
non-Markovian dissipative dynamics, likewise the Markovian case ,
vanishes only at discrete instants. This point agrees with the
results presented in \cite{cavalcanti}, where the authors show
that the  states with zero QD form a set of measure zero.

In the case where the two-qubits interact with the same
environment the entanglement dynamics presents two regimes
\cite{mazzola}: for $\alpha^2\gtrsim 1/4$  damped oscillations of
entanglement are observed [see Fig.3d, dashed line], and for
$\alpha^2\lesssim 1/4$ finite time intervals of complete
disentanglement are followed by entanglement revivals [see
Fig.3b-c, dashed line]. However, while for independent reservoirs
the discord and  entanglement behaviors are similar to each other,
for common environment they behave very differently, as show in
Fig.3a-d. For $\alpha^2=1/2$ both the discord and entanglement
dynamics present the same behavior, but as $\alpha^2$ decreases
the discord exhibits very complicated damped oscillations with
sudden changes more evident. The difference between these two
measures is even more drastic when $\alpha^2=0$, where the
two-qubit state is initially the separable state
$\left|11\right\rangle$. In this case the interaction between the
qubits mediated by a common reservoir does not lead to the
generation of entanglement between them \cite{mazzola}. On the
other hand, the reservoir-mediated interaction leads to the
generation of nonclassical correlations as exposed in Fig. (3a),
showing a ``sudden birth'' of discord but not of entanglement.
Moreover, since the reservoir is initially in the vacuum state,
when the initial state is $\left|00\right\rangle$, obtained from
Eq. (\ref{istate}) with $\alpha^2=1$, no correlation is created
since the composite system is in its ground state. It is also
interesting to note in Fig.3b-c that the decrease of entanglement
is accompanied by the increase of discord in some regions.

In Fig.4 we plotted the discord (solid line) and the EoF (dashed
line) as a function of the scaled time $\gamma_0t$ for a common
reservoir with (a) $\lambda=0.1\gamma_0$, (b) $\lambda=\gamma_0$,
and (c) $\lambda=10\gamma_0$. The two-qubits are initially in the
state (\ref{istate}) with $\alpha^2=1/3$. These results show that
the entanglement decay almost exponentially for
$\lambda=10\gamma_0$, as in the case of Markovian reservoirs. It
is expected because in this regime $\lambda>2\gamma_0$ and
therefore the qubit-reservoir coupling is weak. In this same weak
coupling regime the discord has a similar behavior and we note
that the number of points where the sudden change occur also tends
to decrease. Besides, as the effective coupling between the two
qubits is due to the action of the common reservoir, the quantum
correlations created tends to decrease.

It is worth mentioning that results very similar to ours, for the
case of independent environments, have been numerically studied in
\cite{wang}.

\section{CONCLUSION}
We have studied the quantum discord dynamics of two-qubits coupled
to common and independent non-Markovian environments. We have used
the exactly solvable damped Jaynes-Cummings model for zero
temperature environments. We have observed that even when the
entanglement suddenly disappears and reappears after finite time
intervals, the quantum discord vanishes only at discrete  times.
For a common environment we have observed what is called the
sudden change phenomenon. Actually, the quantum discord between the
qubits suddenly changes depending on the maximization process of
the amount of classical correlations between them. This fact
indicates that this phenomenon could be universal which means
that, for  general initial conditions and interaction
Hamiltonians, we expect that the POVM that maximizes the classical
correlations would abruptly change from one time interval to the
other. This point will be studied in a future work where more
general Hamiltonians and initial conditions will be considered.
Furthermore, we have observed that, in the case of common
environments, the very different behavior of discord and
entanglement can arise even for initially separated states.
Finally, we have also noticed that, even without entanglement, the
correlations introduced by the environment are transferred to the
two qubits producing a finite quantum discord.

\section{Acknowledgments}
We wish to thank the partial financial support from
the Funda{\c c}{\~a}o de Amparo {\`a} Pesquisa de S{\~a}o Paulo (FAPESP) and Conselho Nacional de Desenvolvimento Cient{\'\i}fico Tecnol{\'o}gico (CNPq). We thank Adriana Auyuanet for pointing out a mistake in an early draft. AOC also acknowledges his participation as a member of the Instituto Nacional de Ci{\^e}ncia e Tecnologia em
Informa\c{c}{\~a}ao Qu{\^a}ntica (INCT-IQ).



\begin{thebibliography}{99}
\bibitem{nielsen} M. A. Nielsen and I. L. Chuang, \textit{Quantum
Computation and Quantum Information} (Cambridge University Press,
Cambridge, England, 2000).

\bibitem{zurek} H. Ollivier and W. H. Zurek, Phys. Rev. Lett. \textbf{88},
017901 (2001).


\bibitem{vedral} L. Henderson and V. Vedral, J. Phys. A \textbf{34},
6899 (2001); V. Vedral, Phys. Rev. Lett \textbf{90}, 050401 (2003).

\bibitem{luo} S. Luo, Phys. Rev. A \textbf{77}, 042303 (2008).

\bibitem{datta} A. Datta, A. Shaji, and C. M. Caves, Phys. Rev. Lett. \textbf{100},
050502 (2008).

\bibitem{white} B. P. Lanyon, M. Barbieri, M. P. Almeida, and A. G. White, Phys. Rev. Lett. \textbf{101},
200501 (2008).

\bibitem{phase transition} R. Dillenschneider, Phys. Rev. B \textbf{78},
224413 (2008); M. S. Sarandy, Phys. Rev. A \textbf{80}, 022108 (2009).

\bibitem{Cui} J. Cui and H. Fan , arXiv:0904.2703v1 (2009).

\bibitem{lidar} C. A. Rodriguez-Rosario \textit{et al.}, J. Phys. A: Math.
Theor. \textbf{41}, 205301 (2008); A. Shabani and D. A. Lidar, Phys. Rev.
Lett. \textbf{102}, 100402 (2009).

\bibitem{werlang} T. Werlang, S. Souza, F. F. Fanchini, and C. J. VillasBoas, Phys. Rev. A \textbf{80}, 024103 (2009).

\bibitem{jonas} J. Maziero, L. C. Celeri, R. M. Serra, and V. Vedral, Phys. Rev. A \textbf{80}, 044102 (2009).

\bibitem{cavalcanti} A. Ferraro \textit{et al.}, arXiv:0908.3157v2 (2009).

\bibitem{breuer} H.-P. Breuer and F. Petruccione, \textit{The Theory
of Open Quantum Systems} (Oxford University Press, Oxford, New York,
2002).

\bibitem{eberly} K. Zyczkowski, P. Horodecki, M. Horodecki, and R. Horodecki, Phys. Rev. A \textbf{65}, 012101 (2001); L. {Di\'{o}si}, Lec. Notes Phys. \textbf{622}, 157 (2003); P. J. Dodd and J. J. Halliwell, Phys. Rev. A \textbf{69}, 052105 (2004).

\bibitem{yufirst}  T. Yu and J. H. Eberly, Phys. Rev. Lett. \textbf{93}, 140404 (2004).

\bibitem{citeESD} T. Yu and J. H. Eberly, Phys. Rev. Lett. \textbf{93},
140404 (2004); M. F. Santos, P. Milman, L. Davidovich, and N. Zagury, Phys. Rev. A \textbf{73},
040305(R) (2006); M. P. Almeida \textit{et al.}, Science \textbf{316}, 579
(2007); L. Aolita, R. Chaves, D. Cavalcanti, A. Acin, and L. Davidovich, Phys. Rev. Lett. \textbf{100}, 080501
(2008); A. Salles \textit{et al.}, Phys. Rev. A \textbf{78}, 022322 (2008); C. E. L{\'o}pez, G. Romero, F. Lastra, E. Solano, and J. C. Retamal, Phys. Rev. Lett. \textbf{101}, 080503 (2008).

\bibitem{bellomo07} B. Bellomo, R. Lo Franco, and G. Compagno, Phys.
Rev. Lett. \textbf{99}, 160502 (2007);

\bibitem{ficek} Z. Ficek and R. Tan{\'a}s, Phys. Rev. A \textbf{74}, 024304 (2006).
\bibitem{jppaz} J. P. Paz and A. J. Roncaglia, Phys. Rev. Lett. \textbf{100}, 220401 (2008).
\bibitem{mazzola} L. Mazzola, S. Maniscalco, J. Piilo, K. A. Suominen, and B. M. Garraway, Phys. Rev. A \textbf{79}, 042302 (2009).

\bibitem{nonM} J. Dajka, M. Mierzejewski, and J. Luczka, Phys. Rev. A \textbf{77}, 042316 (2008 ); F. Q. Wang, Z. M. Zhang, and R. S. Liang, Phys. Rev. A \textbf{78}, 062318 (2008); K. Shiokawa, Phys. Rev. A \textbf{79}, 012308 (2009).

\bibitem{ham} S. Hamieh, R. Kobes, and H. Zaraket, Phys. Rev. A \textbf{70}, 052325 (2004).

\bibitem{mazzola01} L. Mazzola, S. Maniscalco, J. Piilo, and K. A. Suominen, arXiv:0904.2857v1 (2009).

\bibitem{bayes} T. M. Cover and J. A. Thomas, \textit{Elements of Information
Theory} (Wiley-Interscience, New York, 2006).

\bibitem{3level} B. J. Dalton and B. M. Garraway, Phys. Rev. A \textbf{68}, 033809 (2003); B. J. Dalton, S. M. Barnett, and B. M. Garraway, Phys. Rev. A \textbf{64}, 053813 (2001).

\bibitem{pseudomode} B. M. Garraway, Phys. Rev. A \textbf{55}, 2290 (1997).

\bibitem{eof} C. H. Bennett, D. P. DiVincenzo, J. A. Smolin, and W. K. Wootters, Phys. Rev. A \textbf{54}, 3824 (1996).

\bibitem{concurrence} W. K. Wootters, Phys. Rev. Lett. \textbf{80}, 2245 (1998).

\bibitem{wang} B. Wang, Z. Xu, Z. Chen, and M. Feng, arXiv:0911.1845v1 (2009).
\end{thebibliography}
\end{document}